# Informed peer review for publication assessments: Are improved impact measures worth the hassle?[1]

**Authors:** Giovanni Abramo[1], Ciriaco Andrea D'Angelo[2,*], Giovanni Felici[3]

**Affiliations:**

[1] Laboratory for Studies in Research Evaluation, Institute for System Analysis and Computer Science (IASI-CNR). National Research Council, Rome, Italy – giovanni.abramo@uniroma2.it – ORCID: 0000-0003-0731-3635

[2] University of Rome "Tor Vergata", Dept of Engineering and Management, Rome, Italy – dangelo@dii.uniroma2.it – ORCID: 0000-0002-6977-6611

[3] Institute for System Analysis and Computer Science (IASI-CNR). National Research Council, Rome, Italy - giovanni.felici@iasi.cnr.it – ORCID: 0000-0003-0544-5407

*\* corresponding author*

**Abstract**
In this work we ask whether and to what extent applying a predictor of publications' impact better than early citations, has an effect on the assessment of research performance of individual scientists. Specifically, we measure the total impact of Italian professors in the sciences and economics in a period of time, valuing their publications first by early citations and then by a weighted combination of early citations and impact factor of the hosting journal. As expected, scores and ranks by the two indicators show a very strong correlation, but there occur also significant shifts in many fields, mainly in Economics and statistics, and Mathematics and computer science. The higher the share of uncited professors in a field and the shorter the citation time window, the more recommendable the recourse to the above combination.

**Keywords**
*Research evaluation; citation time window; bibliometrics; Italy*

---

# 1. Introduction

Evaluative scientometrics is mainly aimed at measuring and comparing research performance of entities. In general, a research entity is said to perform better than another if, all production factors being equal, its total output has higher impact. The question then is how to measure the impact of output. Citation-based indicators are more apt to assess scholarly impact than social impact, although it is reasonable to expect that a certain correlation between scholarly and social impact exists (Abramo, 2018).

As far as scholarly impact is concerned, three approaches are available to assess the impact of publications: by human judgment (peer review), or through the use of citation-based indicators (bibliometrics), or by drawing on both whereby bibliometrics informs peer-review judgment (informed peer review).

The axiom underlying citation-based indicators is that when a publication is cited, it has contributed to (has had an impact on) the new knowledge encoded in the citing publications - normative theory (Kaplan, 1965; Merton, 1973; Bornmann & Daniel, 2008). There are strong distinctions and objections to the above axiom argued by the social constructivism school, holding that that citing to give credit is the exception, while persuasion is the major motivation for citing (Mulkay, 1976; Bloor, 1976; Gilbert, 1977; Latour, 1987; Brooks, 1985, 1986; MacRoberts & MacRoberts, 1984, 1987, 1988, 1989a, 1989b, 1996, 2018; Teplitskiy, Dueder, Menietti, & Lakhani, 2019).

Although scientometricians for short say that they "measure" scholarly impact, what they actually do is "predicting" impact. The reason is that to serve its purpose, any research assessment aimed at informing policy and management decisions cannot wait for the publications life-cycle to be completed (i.e. the publications stop being cited), which may take decades (van Raan, 2004; Teixeira, Vieira, & Abreu, 2017; Song, Situ, Zhu, & Lei, 2018).

As a consequence, scientometricians count early citations, not overall. The question then is how long should the citation time window be, in order for the early citations to be considered an accurate and robust proxy of overall scholarly impact. The longer the citation time window, the more accurate the prediction. In the end, the answer is subjective, because of the embedded tradeoff: the appropriate choice of citation time window is a compromise between the two objectives of accuracy and timeliness in measurement, and the relative solutions differ from one discipline to another. The topic has been extensively examined in literature (Rousseau, 1988; Glänzel, Schlemmer, & Thijs, 2003; Adams, 2005; Stringer, Sales-Pardo, & Nunes Amaral, 2008; Abramo, Cicero, & D'Angelo, 2011; Nederhof, Van Leeuwen, & Clancy, 2012 Wang, 2013; Onodera, 2016).

Most studies in evaluative scientometrics focus on providing new creative solutions to the problem of how to best support the measurement of research performance. An extraordinary number of performance indicators continue to be proposed. It suffices to say that in the recent *17th International Society of Scientometrics and Informetrics Conference - (ISSI 2019),* a special plenary session and five parallel sessions, including 25 contributions altogether (leaving aside poster presentations), were devoted to "novel bibliometric indicators".

Much fewer studies have tackled the problem of how to improve the impact prediction power of early citations, given inevitable citation short time windows. A number of scholars proposed to combine citation counts with other independent variables related to the publication. Whichever the combination, there is a common awareness that it cannot be the same across disciplines, because the citation accumulation speed and distribution curves vary across disciplines (Garfield, 1972; Mingers, 2008; Wang, 2013; Baumgartner & Leydesdorff, 2014).

It has been shown that in mathematics (and with weaker evidence in biology and earth sciences), for two-years or less citation windows, the journal's two-year impact factor (*IF*) predicts better than early citations long-term impact (Abramo, D'Angelo, & Di Costa, 2010). In all disciplines of the sciences but mathematics, for citation windows of zero or one year only, a combination of *IF* and citations was recommended (Levitt & Thelwall, 2011; Bornmann, Leydesdorff, & Wang 2014). The same seems to be valid in the social sciences as well (Stern, 2014). A model based on *IF* and citations to predict long-term citations was proposed by Stegehuis, Litvak, and Waltman (2015). The weighted combination of citations and journal metric percentiles adopted in the Italian research assessment exercise, VQR 2011-2014 (Anfossi, Ciolfi, Costa, Parisi, & Benedetto, 2016), proved to be a worse predictor of future impact than citations only (Abramo and D'Angelo, 2016).

To provide practitioners and decision makers with a better predictor of overall impact, and awareness of how the predicting power varies with the citation time window, Abramo, D'Angelo, and Felici (2019) made available, in each of the 170 subject categories (SCs) in the sciences and economics, with more than 100 Italian 2004-2006 publications: i) the weighted combinations of two-year *IF* and citations, as a function of the citation time window, which best predict overall impact; and ii) the predictive power of each combination.

It emerged that the IF has a non-negligible role only with very short citation time windows (0 to 2 years); for longer ones, the weight of early citations is dominating and the IF is not informative in explaining the difference between long-term and short-term citations.

The calibration of the weights by citation time window and SC, and the measurement of the impact indicator is not so straightforward as the simple measurement of normalized citations.

In this study, we want to find out whether all this hassle about improving the predicting power of early citations is worthwhile. We ask whether and to what extent applying a predictor of overall impact more accurate than early citations, has an effect on the research performance ranks of individuals. In this specific case, as a performance indicator we recur to total impact of individuals. This indicator is particularly appropriate when one needs to identify the top experts in a particular field, for consultancy work or the like. Counting on an authors' name disambiguation algorithm of Italian academics, we measure the total impact of Italian professors (assistant, associate, full) in the sciences and in economics in a period of time, valuing their publications first by the early citations and then by the weighted citation-*IF* combination provided by Abramo, D'Angelo, and Felici (2019). At this point, we can analyze the extent of variations in rank of individuals, in each discipline and field where they are classified.[2]

The rest of the manuscript is organized as follows. In Section 2, we present the data and method. In Section 3, we report the comparison of the rankings by the two methods of valuing overall impact, at field and discipline level. The discussion of results in Section 4 will conclude the work.

---

[2] To accomplish the assignment, first we need integrate the IF-citation combinations calculated in the above said 170 SCs, with those in the other SCs where the population under observation publishes.

## 2. Data and methods

For the purpose of this study, we are interested in how a different measure of impact affects the ranking of Italian professors by total impact, in the period of 2015-2017.

Data on the faculty at each university were extracted from the database on Italian university personnel, maintained by the Ministry of Universities and Research, MUR. For each professor this database provides information on their gender, affiliation, field classification and academic rank, at the end of each year.[3] In the Italian university system all academics are classified in one and only one field, named scientific disciplinary sector (SDS), 370 in all. SDSs are grouped into disciplines, named university disciplinary areas (UDAs), 14 in all.

Data on output and relevant citations are extracted from the Italian Observatory of Public Research, a database developed and maintained by Abramo and D'Angelo, and derived under license from the Clarivate Analytics Web of Science Core Collection (WoS). Beginning from the raw data of the WoS, and applying a complex algorithm to reconcile the author's affiliation and disambiguation of the true identity of the authors, each publication (article, letter, review and conference proceeding) is attributed to the university professor that produced it.[4] Thanks to this algorithm, we can produce rankings by total impact at the individual level, on a national scale. Based on the value of *total impact* we obtain a ranking list expressed on a percentile scale of 0-100 (worst to best) of all Italian academics of the same academic rank and SDS.

We limit our field of analysis to the sciences and economics, where the WoS coverage is acceptable for bibliometric assessment. The dataset thus formed consists of 38,456 professors from 11 UDAs (mathematics and computer sciences, physics, chemistry, earth sciences, biology, medicine, agricultural and veterinary sciences, civil engineering, industrial and information engineering, psychology, economics and statistics) and 218 SDSs, as shown in Table 1. 9.3% of professors are unproductive (0 publications), and as a consequence their scores remain unchanged by the two indicators, but not necessarily their ranks. In fact, the scores and ranks of uncited productive professors (4.2% in all) will change (because *IF* is always above 0). Measuring the latter's impact by citations only, their score (0) and rank would be the same as for unproductive professors. It would not, when measured by the weighted combination of normalized citations and *IF*.

As for impact, we measure it in two ways: one way values publications by early citations only; and the other by the weighted combinations of citations and IF,[5] as a function of the citation time window and field of research, which best predict future impact (Abramo, D'Angelo, & Felici, 2019). Because citation behavior varies across fields, we standardize the citations for each publication with respect to the average of the distribution of citations for all publications indexed in the same year and the same SC.[6] We apply the same procedure to the *IF*. Furthermore, research projects frequently involve a team of scientists, which is registered in the co-authorship of publications. In this case, we account for the fractional contributions of scientists to outputs, which is sometimes further signaled by the position of the authors in the list of authors.

---

[3] http://cercauniversita.cineca.it/php5/docenti/cerca.php, last accessed on 31 March, 2020.
[4] The harmonic average of precision and recall (F-measure) of authorships, as disambiguated by the algorithm, is around 97% (2% margin of error, 98% confidence interval).
[5] The journal IF refers to the year of publication.
[6] Abramo, Cicero, and D'Angelo (2012) demonstrated that the average of the distribution of citations received for all cited publications of the same year and SC is the best-performing scaling factor.

*Table 1: Dataset of the analysis. Italian professors holding formal faculty roles, for at least two years over the 2015-2017 period, by UDA and academic rank.*

| UDA* | N. of SDSs | Total professors | Unproductive | Uncited productive |
|---|---|---|---|---|
| 1 | 10 | 3019 | 380 (12.6%) | 227 (7.5%) |
| 2 | 8 | 2146 | 103 (4.8%) | 42 (2.0%) |
| 3 | 11 | 2815 | 59 (2.1%) | 23 (0.8%) |
| 4 | 12 | 1010 | 50 (5.0%) | 11 (1.1%) |
| 5 | 19 | 4630 | 184 (4.0%) | 53 (1.1%) |
| 6 | 50 | 9159 | 748 (8.2%) | 231 (2.5%) |
| 7 | 30 | 2948 | 190 (6.4%) | 76 (2.6%) |
| 8 | 9 | 1500 | 129 (8.6%) | 63 (4.2%) |
| 9 | 42 | 5290 | 246 (4.7%) | 169 (3.2%) |
| 10 | 10 | 1402 | 168 (12.0%) | 68 (4.9%) |
| 11 | 17 | 4537 | 1312 (28.9%) | 635 (14.0%) |
| Total | 218 | 38456 | 3569 (9.3%) | 1598 (4.2%) |

\* *1, Mathematics and computer science; 2, Physics; 3, Chemistry; 4, Earth sciences; 5, Biology; 6, Medicine; 7, Agricultural and veterinary sciences; 8, Civil engineering; 9, Industrial and information engineering; 10, Psychology; 11, Economics and statistics.*

The yearly total impact of a professor, termed *TI*, is then defined as:

$$TI = \frac{1}{t}\sum_{i=1}^{N} c_i f_i$$

[2]

where:
$t$ = number of years on staff of professor during the observation period
$N$ = number of publications by the professor in period under observation;
$c_i$ = alternatively: i) citations received by publication *i* normalized to the average of distribution of citations received for all cited publications in same year and SC of publication *i*; ii) weighted combination of normalized citations and normalized *IF* of the hosting journal, whereby weights differ across citation time windows and SCs, as in Abramo, D'Angelo, and Felici (2019);
$f_i$ = fractional contribution of professor to publication *i*.

The fractional contribution equals the inverse of the number of authors in those fields where the practice is to place the authors in simple alphabetical order but assumes different weights in other cases. For the life sciences, widespread practice in Italy is for the authors to indicate the various contributions to the published research by the order of the names in the listing of the authors. For the life sciences, we give then different weights to each co-author according to their position in the list of authors and the character of the co-authorship (intra-mural or extra-mural).[7]

For reasons of significance, the analysis is limited to those professors who held formal faculty roles, for at least two years over the 2015-2017 period.

Citations are observed at 31 December 2018, implying citation time windows ranging from one to four years.

---

[7] If the first and last authors belong to the same university, 40% of the citation is attributed to each of them, the remaining 20% is divided among all other authors. If the first two and last two authors belong to different universities, 30% of the citation is attributed to the first and last authors, 15% of the citation is attributed to the second and last authors but one, the remaining 10% is divided among all others. The weightings were assigned following advice from senior Italian professors in the life sciences. The values could be changed to suit different practices in other national contexts.

## 3. Results

In the following, we present the score and rank of performance by total impact of Italian professors, by SDS and UDA, as measured respectively by:
- Early citations ($TI_C$);
- The weighted combination of citations and *IF* of the hosting journal ($TI_{WC}$).

As already said, no variations will occur for professors with no publications in the period under observation. We expect instead significant variations in score and rank for professors with uncited publications. In fact, while $TI_C$ is nil, $TI_{WC}$ is going to be above 0.

As an example, Table 2 shows the scores and ranks by $TI_C$ and $TI_{WC}$, for the 26 Italian professors in the SDS Aerospace propulsion. The score variation is nil for the two unproductive professors at the bottom of the list, while it is maximum for the uncited productive professors (ID 49113 and 2592). Twelve professors experience no shift, among them the top five in ranking. Few pairs swap positions e.g. ID 78162 and ID 49106. The maximum shift is 3 positions.

*Table 2: Ranking lists by total impact ($TI_c$ and $TI_{WC}$) of Italian professors in the SDS Aerospace propulsion*

| ID | $TI_C$ score | rank | percentile | $TI_{WC}$ score | rank | percentile | Δ score | Δ rank | |
|---|---|---|---|---|---|---|---|---|---|
| 10712 | 2.703 | 1 | 100 | 3.360 | 1 | 100.0 | 24.3% | 0 | = |
| 49114 | 0.824 | 2 | 96 | 1.268 | 2 | 96.0 | 53.9% | 0 | = |
| 49109 | 0.773 | 3 | 92 | 0.906 | 3 | 92.0 | 17.2% | 0 | = |
| 4045 | 0.666 | 4 | 88 | 0.853 | 4 | 88.0 | 28.2% | 0 | = |
| 2590 | 0.633 | 5 | 84 | 0.759 | 5 | 84.0 | 19.8% | 0 | = |
| 78162 | 0.548 | 6 | 80 | 0.698 | 7 | 76.0 | 27.5% | 1 | ↓ |
| 49106 | 0.504 | 7 | 76 | 0.731 | 6 | 80.0 | 44.9% | 1 | ↑ |
| 4047 | 0.365 | 8 | 72 | 0.489 | 8 | 72.0 | 34.0% | 0 | = |
| 37761 | 0.240 | 9 | 68 | 0.383 | 10 | 64.0 | 59.4% | 1 | ↓ |
| 4044 | 0.224 | 10 | 64 | 0.479 | 9 | 68.0 | 113.7% | 1 | ↑ |
| 2597 | 0.211 | 11 | 60 | 0.340 | 11 | 60.0 | 61.4% | 0 | = |
| 5463 | 0.191 | 12 | 56 | 0.287 | 12 | 56.0 | 50.7% | 0 | = |
| 49118 | 0.183 | 13 | 52 | 0.268 | 13 | 52.0 | 46.7% | 0 | = |
| 49115 | 0.105 | 14 | 48 | 0.132 | 14 | 48.0 | 26.1% | 0 | = |
| 49117 | 0.074 | 15 | 44 | 0.085 | 18 | 32.0 | 15.0% | 3 | ↓ |
| 78159 | 0.069 | 16 | 40 | 0.103 | 15 | 44.0 | 48.6% | 1 | ↑ |
| 2595 | 0.059 | 17 | 36 | 0.085 | 17 | 36.0 | 43.3% | 0 | = |
| 4046 | 0.059 | 17 | 36 | 0.072 | 19 | 28.0 | 21.3% | 2 | ↓ |
| 4048 | 0.047 | 19 | 28 | 0.099 | 16 | 40.0 | 110.6% | 3 | ↑ |
| 49111 | 0.036 | 20 | 24 | 0.038 | 21 | 20.0 | 5.6% | 1 | ↓ |
| 2589 | 0.024 | 21 | 20 | 0.025 | 22 | 16.0 | 5.6% | 1 | ↓ |
| 87212 | 0.020 | 22 | 16 | 0.040 | 20 | 24.0 | 97.5% | 2 | ↑ |
| 49113 | 0.000 | 23 | 0 | 0.012 | 23 | 12.0 | ∞ | 0 | = |
| 2592 | 0.000 | 23 | 0 | 0.004 | 24 | 8.0 | ∞ | 1 | ↓ |
| 2599 | 0.000 | 23 | 0 | 0.000 | 25 | 0.0 | n.a. | 2 | ↓ |
| 40946 | 0.000 | 23 | 0 | 0.000 | 25 | 0.0 | n.a. | 2 | ↓ |

The SDS Industrial chemistry consists of 114 professors, mostly productive and cited. Figure 1 shows the dispersion of their impact. The very strong correlation of scores (Pearson $\rho = 0.999$) and ranks (Spearman $\rho = 0.998$) by $TI_c$ and $TI_{WC}$ are as expected.

*Figure 1: Score dispersion by $TI_C$ and $TI_{WC}$ of the 114 Italian professors in the SDS Industrial chemistry*

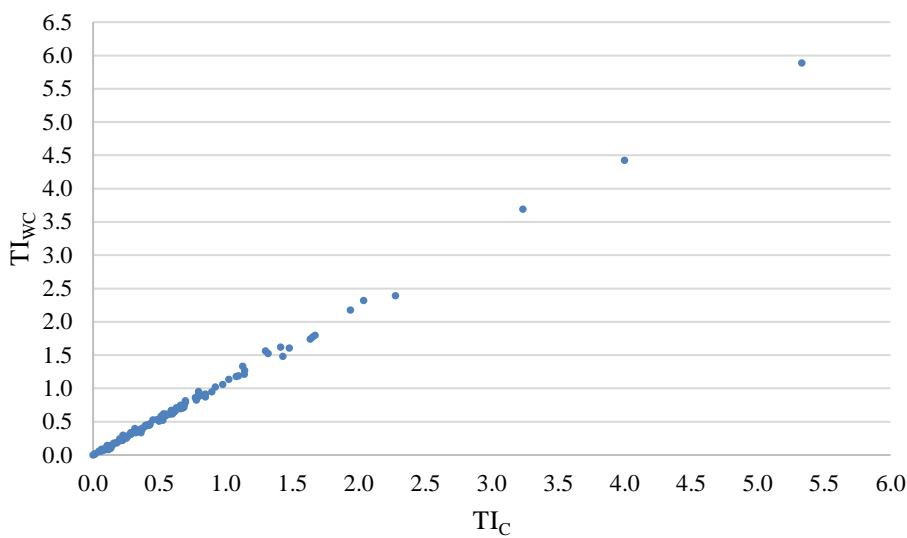

Higher dispersion (Figure 2) occurs instead for the 73 professors in the SDS Complementary mathematics, whereby about two thirds (50) of professors present nil $TI_C$, and 20% (15) while productive are uncited ($TI_C$ above 0). As a matter of fact, noticeable shifts in relative scores occur for high performers too (right-top side of the diagram), notwithstanding a very strong score correlation (Pearson $\rho = 0.988$). The ability of $TI_{WC}$ to discriminate the impact of uncited publications, and therefore the relevant performance of uncited professors, explains the lower rank correlation (Spearman $\rho = 0.915$). Although variations in score are not that noticeable, those in rank are. To better show that, Figure 3 reports the share of professors experiencing a rank shift in both SDSs. In Complementary mathematics, above 60% of professors do not change rank (50% could not, as unproductive). The remaining 40% though present shifts, which are in some cases quite noticeable: five professors improve their rank by no less than 10 positions. Rank shifts are less evident in Industrial chemistry: the average shift is 1.47 positions, as compared to 1.89 Complementary mathematics. In the former SDS, because of the lower number of unproductive professors, shifts concern a higher share of the population, namely 70%.

For a better appreciation of the rank variations in the whole SDS spectrum Figure 4 shows the box plots of the average percentile shifts in the SDSs of each UDA, while Table 3 presents some relevant descriptive statistics.

Economics and statistics is the UDA with the highest average percentile shift (6.5), the highest dispersion among SDSs (3.6 standard deviation), and the widest range of percentile shift, from 1.4 of SECS-P/13 (Commodity science) to 15.9 of SECS-P/04 (History of economic thought). It is followed by Mathematics and computer science, whose range of variation of the percentile shift is between 2.0 of MAT/08 (Numerical analysis) and 12.6 of MAT/04 (Complementary mathematics). On the contrary, UDAs 4 (Earth sciences) and 5 (Biology) show the lowest dispersion among SDSs (0.3 standard deviation) and quite low average percentile shifts. In UDA Medicine, a peculiar case occurs: in SDS MED/47 (Nursing and midwifery) the two ranking lists are exactly the same. The same occurs also in two SDSs of Industrial and information engineering: ING-IND/29 (Raw materials engineering) and ING-IND/30 (Hydrocarburants and fluids of the subsoil). In general, in 17 out of 218 SDSs, the average percentile shift is never below five percentiles.

*Figure 2: Score dispersion by $TI_c$ and $TI_{WC}$ of the 73 Italian professors in the SDS Complementary mathematics*

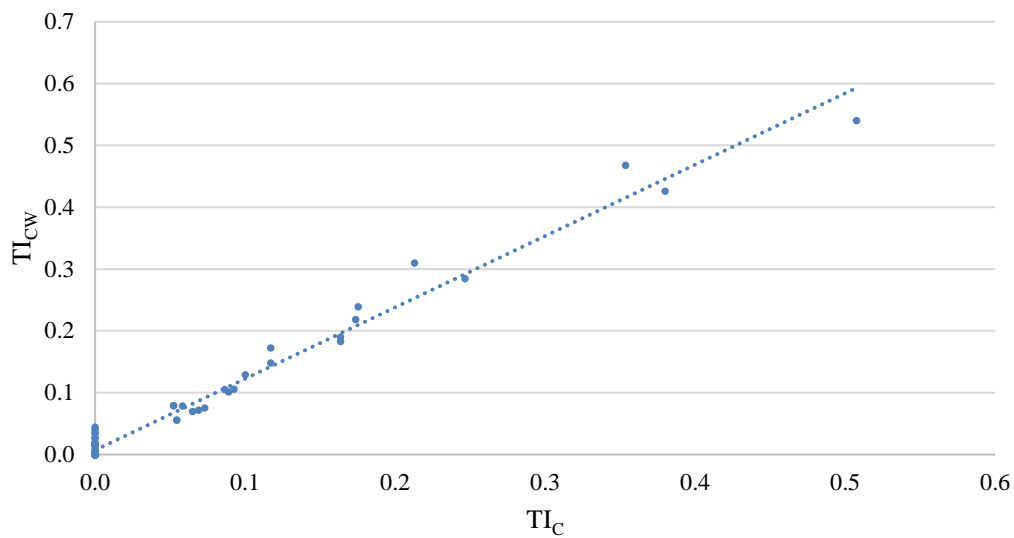

*Figure 3: Share of professors experiencing a rank shift in the SDSs Industrial chemistry (CHEM/04) and Complementary mathematics (MATH/04)*

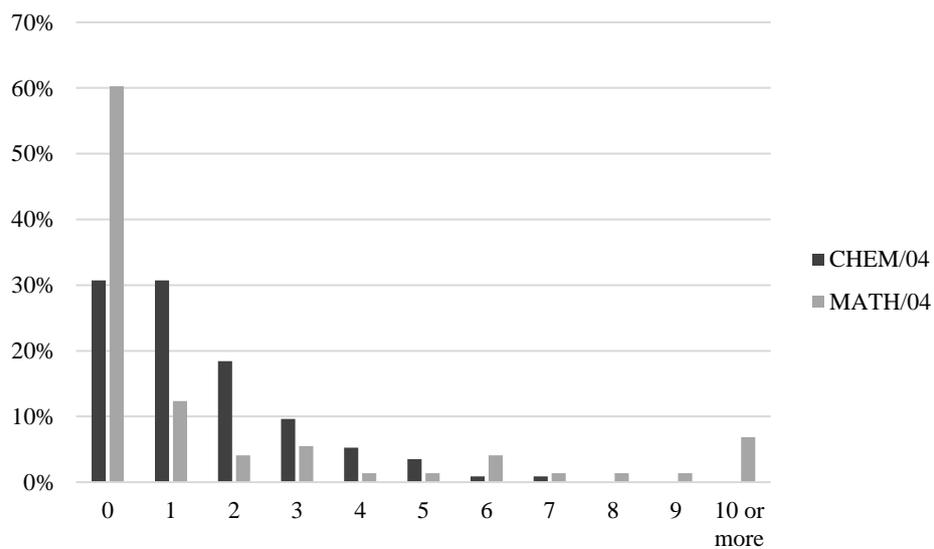

*Figure 4: Box plot of average percentile shifts in the SDSs of each UDA*

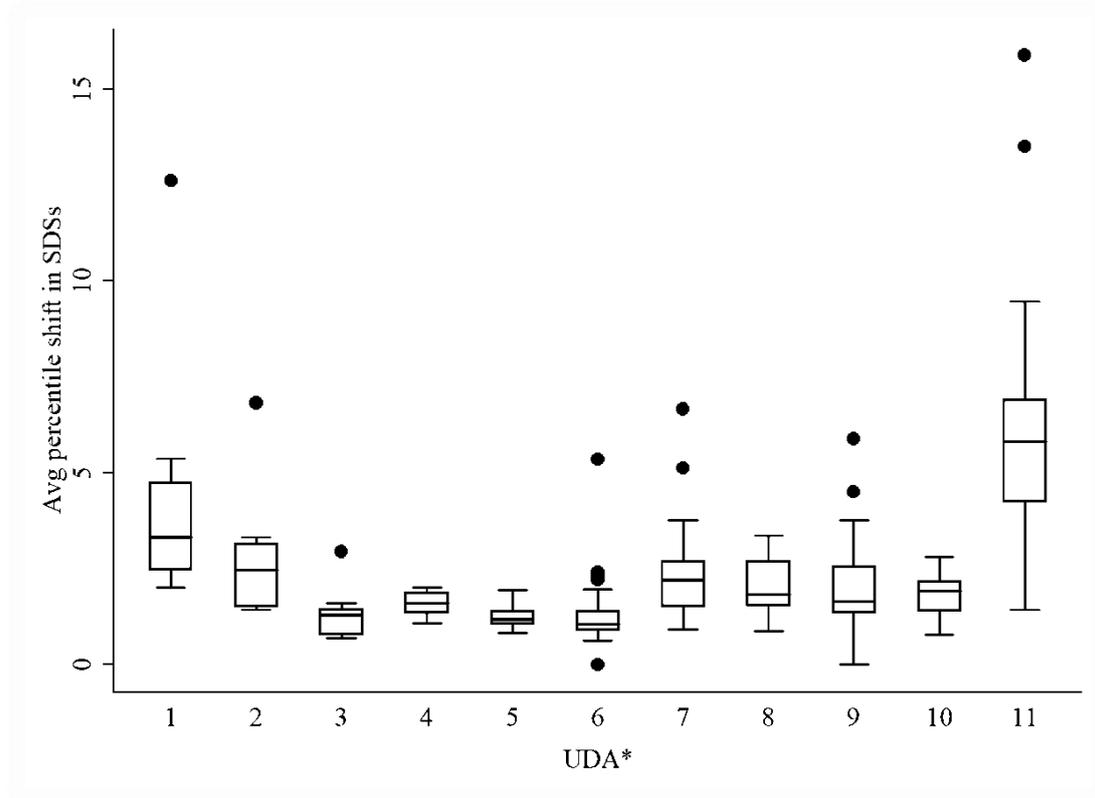

* 1, Mathematics and computer science; 2, Physics; 3, Chemistry; 4, Earth sciences; 5, Biology; 6, Medicine; 7, Agricultural and veterinary sciences; 8, Civil engineering; 9, Industrial and information engineering; 10, Psychology; 11, Economics and statistics.

*Table 3: Descriptive statistics of percentile shifts in the SDSs of each UDA*

| UDA | Min | Max | Avg | St. dev. |
|---|---|---|---|---|
| 1 | 2.0 (MAT/08) | 12.6 (MAT/04) | 4.3 | 3.0 |
| 2 | 1.4 (FIS/03) | 6.8 (FIS/08) | 2.8 | 1.7 |
| 3 | 0.7 (CHIM/09) | 3.0 (CHIM/12) | 1.3 | 0.6 |
| 4 | 1.1 (GEO/07) | 2.0 (GEO/10) | 1.6 | 0.3 |
| 5 | 0.8 (BIO/14) | 1.9 (BIO/05) | 1.3 | 0.3 |
| 6 | 0 (MED/47) | 5.3 (MED/02) | 1.3 | 0.7 |
| 7 | 0.9 (AGR/16) | 6.7 (AGR/06) | 2.3 | 1.2 |
| 8 | 0.9 (ICAR/03) | 3.4 (ICAR/06) | 2.0 | 0.7 |
| 9 | 0 (ING-IND/29; ING-IND/30) | 5.9 (ING-IND/02) | 1.9 | 1.1 |
| 10 | 0.8 (M-EDF/02) | 2.8 (M-PSI/07) | 1.8 | 0.6 |
| 11 | 1.4 (SECS-P/13) | 15.9 (SECS-P/04) | 6.5 | 3.6 |

* AGR/06, Wood technology and forestry operations; AGR/16, Agricultural Microbiology; BIO/05, Zoology; BIO/14, Pharmacology; CHIM/09, Pharmaceutical and technological applications of chemistry; CHIM/12, Chemistry for the environment and for cultural heritage; FIS/03, Physics of matter; FIS/08, Didactics and history of physics; GEO/07, Petrology and petrography; GEO/10, Solid Earth geophysics; ICAR/03, Sanitary and environmental engineering; ICAR/06, Topography and cartography; ING-IND/02, Ship structures and marine engineering; ING-IND/29, Engineering of raw materials; ING-IND/30, Hydrocarbons and underground fluids; MAT/04, Mathematics education and history of mathematics; MAT/08, Numerical analysis; MED/02, Medical history; MED/47, Midwifery; M-EDF/02, Methods and teaching of sports activities; M-PSI/07, Dynamic psychology; SECS-P/04, History of economic thought; SECS-P/13, Commodity science.

In general, the correlation between $TI_C$ and $TI_{WC}$ is very strong. Table 4 presents some descriptive statistics of both Pearson ρ (score) and Spearman ρ (rank) for the SDSs of each UDA. As for the scores, the minimum correlation (0.957) occurs in an SDS of Medicine (MED/02 - History of medicine). As for the ranks, it occurs (0.884) in an SDS of Economics and statistics, SECS-P/04 (History of economic thought), outstanding also for the maximum average percentile shift among all SDSs (Table 3). It is a relatively small SDS, 35 professors in all, two thirds of which with nil $TI_C$.

*Table 4: Descriptive statistics of correlation coefficients for $TI_C$ and $TI_{WC}$ in the SDSs of each UDA*

| UDA* | Pearson correlation | | | | Spearman correlation | | | |
|---|---|---|---|---|---|---|---|---|
| | Min | Max | Avg | St. dev. | Min | Max | Avg | St. dev. |
| 1 | 0.986 | 0.999 | 0.995 | 0.004 | 0.915 | 0.995 | 0.981 | 0.023 |
| 2 | 0.984 | 0.999 | 0.994 | 0.005 | 0.969 | 0.997 | 0.990 | 0.009 |
| 3 | 0.998 | 1 | 0.999 | 0.001 | 0.973 | 0.999 | 0.996 | 0.007 |
| 4 | 0.997 | 1 | 0.999 | 0.001 | 0.994 | 0.998 | 0.996 | 0.001 |
| 5 | 0.998 | 1 | 0.999 | 0.001 | 0.995 | 0.999 | 0.998 | 0.001 |
| 6 | 0.957 | 1 | 0.998 | 0.006 | 0.964 | 1 | 0.997 | 0.005 |
| 7 | 0.981 | 1 | 0.997 | 0.004 | 0.939 | 0.999 | 0.992 | 0.011 |
| 8 | 0.994 | 0.999 | 0.998 | 0.002 | 0.989 | 0.999 | 0.995 | 0.003 |
| 9 | 0.983 | 1 | 0.997 | 0.003 | 0.953 | 1 | 0.994 | 0.008 |
| 10 | 0.995 | 1 | 0.998 | 0.002 | 0.994 | 0.999 | 0.996 | 0.002 |
| 11 | 0.993 | 0.998 | 0.996 | 0.002 | 0.884 | 0.997 | 0.969 | 0.029 |

\* *1, Mathematics and computer science; 2, Physics; 3, Chemistry; 4, Earth sciences; 5, Biology; 6, Medicine; 7, Agricultural and veterinary sciences; 8, Civil engineering; 9, Industrial and information engineering; 10, Psychology; 11, Economics and statistics.*

The rank variations in general appear strongly correlated with the share of productive professors with nil $TI_C$, i.e. with only uncited publications. The correlation between the two variables is shown in Figure 5 (Pearson ρ = 0.791).

*Figure 5: Field dispersion per share of uncited professors and average rank shift by $TI_C$ and $TI_{WC}$*

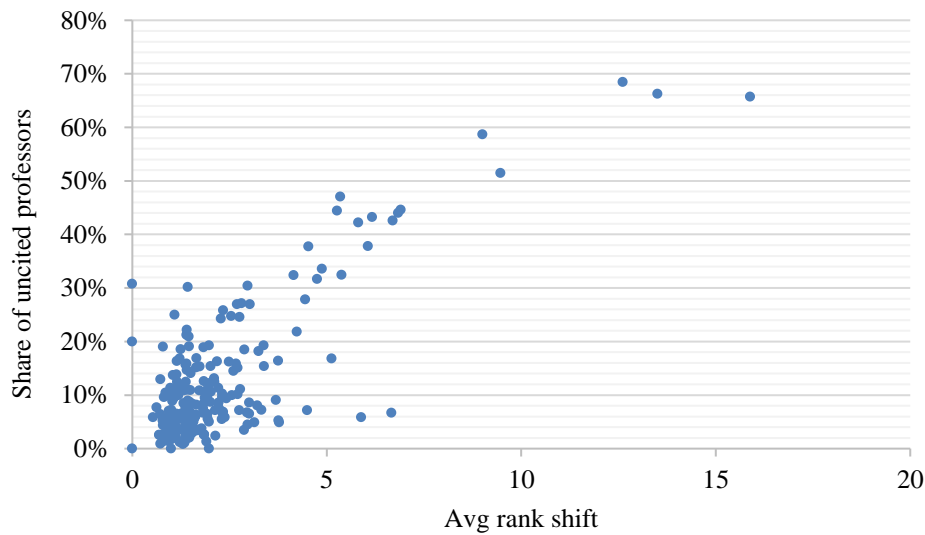

A typical way to report performance is by quartile ranking. We then analyse the performance quartile shifts by the two indicators of impact. Table 5 represents the contingency matrix of the performance quartile by $TI_C$ and $TI_{WC}$ of all 38456 professors of the dataset. Because of the strong correlation between the two indicators, we observe an equally strong concentration of frequencies along the main diagonal: in 93.3% of cases (24.52%+22.90%+20.96%+24.94%), the performance quartile remains unchanged. 0.7% of professors in Q1 by $TI_C$ shift to Q2 by $TI_{WC}$. Alongside, 0.8% of professors in Q4 by $TI_C$ shift to Q3 by $TI_{WC}$.

Table 6 shows the shift distributions by UDA. Economics and statistics presents the highest share of professors (17.1%) shifting quartile, followed by Mathematics (9%). In the remaining UDAs shares range between 4% and 7%. It must be noted that 0.4% of professors (154) experience two quartile shifts, and all but two shift from bottom to above the median. They are mainly in Economics and statistics, and in Mathematics.

*Table 5: Professors' performance quartile distribution as measured by $TI_C$ and $TI_{WC}$*

|  |  | \multicolumn{4}{c}{$TI_{WC}$} | | | |
|---|---|---|---|---|---|
|  |  | I | II | III | IV |
| $TI_C$ | I | 24.52% | 0.70% | 0.00% | 0.00% |
|  | II | 0.69% | 22.90% | 1.03% | 0.00% |
|  | III | 0.00% | 0.96% | 20.96% | 0.84% |
|  | IV | 0.00% | 0.40% | 2.06% | 24.94% |

*Table 6: Distribution of quartile shifts based on $FSS_P$ as measured by $TI_C$ and $TI_{WC}$, by UDA (in brackets, percentage referred to the total UDA staff)*

| UDA* | Shifting quartiles | Shifting two quartiles |
|---|---|---|
| 1 | 271 (9.0%) | 15 (0.50%) |
| 2 | 160 (7.5%) | 0 (0%) |
| 3 | 122 (4.3%) | 1 (0.04%) |
| 4 | 63 (6.2%) | 0 (0%) |
| 5 | 200 (4.3%) | 0 (0%) |
| 6 | 349 (3.8%) | 1 (0.01%) |
| 7 | 197 (6.7%) | 1 (0.03%) |
| 8 | 77 (5.1%) | 0 (0%) |
| 9 | 284 (5.4%) | 0 (0%) |
| 10 | 71 (5.1%) | 0 (0%) |
| 11 | 778 (17.1%) | 136 (3.00%) |
| Total | 2572 (6.7%) | 154 (0.40%) |

\* 1, Mathematics and computer science; 2, Physics; 3, Chemistry; 4, Earth sciences; 5, Biology; 6, Medicine; 7, Agricultural and veterinary sciences; 8, Civil engineering; 9, Industrial and information engineering; 10, Psychology; 11, Economics and statistics.

## 4. Conclusions

Evaluative scientometrics is mainly aimed at measuring and comparing research performance of individuals and organizations. A critical issue in the process is the accurate prediction of scholarly impact of publications, when citation short time windows are allotted. This is often the case, when the evaluation is geared to informed decision making.

A better impact prediction accuracy often involves complex, costly and time-consuming measurements. Pragmatism asks for an analysis of the effects of improved indicators on the performance ranking of the subjects under evaluation. This study follows up the work by the same authors (Abramo, D'Angelo & Felici, 2019), which demonstrated that especially with very short time windows (0 to 2 years) the *IF* can be combined with early citations, as a powerful covariate for predicting long term impact.

Using the outcomes of such inspiring work, i.e. the weighted combinations of *IF* and citations (as a function of the citation time window), which best predict overall impact of single publications in each SC, we have been able to measure the 2015-2017 total impact of all Italian professors in the sciences and economics, and to analyze the entity of variations in performance ranks when using early citations only.

As expected, scores and ranks by the two indicators show a very strong correlation. Nevertheless, in 7% of SDSs, the average shift results never below 5 percentiles, and 15.6 and 12.9 on average in the SDSs, respectively, of Economics and statistics, and Mathematics and computer science.

In terms of quartile shifts, almost 7% of professors undergo them. In Economics and statistics, 3% of professors shift from Q4 to above the median.

It is to be noticed a strong correlation between the rate of shifts in rank and the share of uncited professors in the SDS. The total impact of uncited professors is in fact nil by $TI_C$, but above 0 by $TI_{WC}$. In short, $TI_{WC}$ can better discriminate the performance of professors in the left tail of the distribution. The higher the share of uncited professors in an SDS, the more recommendable the recourse to $TI_{WC}$ is. Furthermore, the shorter the citation time window, the heavier the relative weight of *IF* in predicting the long-term impact. $TI_{WC}$ is then highly recommendable when citation time windows are short and the rate of uncited professors are high.

In the case of national research assessment exercises based on informed peer-review or on bibliometrics only, the weighted combination of normalized citations and *IF* to rank publications might be adopted, as the weights can be made available by the authors, for all SCs and citation time windows up to six years.

Possible future investigations within this stream of research, might concern the effect of the improved indicator of publications' impact on the performance score and rank of research organizations and research units.

**References**

Abramo, G. (2018). Revisiting the scientometric conceptualization of impact and its measurement. *Journal of Informetrics,* 12(3), 590-597.

Abramo, G., Cicero, T., & D'Angelo, C.A. (2011). Assessing the varying level of impact measurement accuracy as a function of the citation window length. *Journal of Informetrics*, 5(4), 659-667.

Abramo, G., Cicero, T., & D'Angelo, C.A. (2012). Revisiting the scaling of citations for research assessment. *Journal of Informetrics*, 6(4), 470-479.

Abramo, G., & D'Angelo, C.A. (2016). Refrain from adopting the combination of citation and journal metrics to grade publications, as used in the Italian national research assessment exercise (VQR 2011-2014). *Scientometrics*, 109(3), 2053-2065.

Abramo, G., D'Angelo, C.A., & Di Costa, F. (2010). Citations versus journal impact factor as proxy of quality: Could the latter ever be preferable? *Scientometrics*, 84(3), 821-833.

Abramo, G., D'Angelo, C.A., & Felici, G. (2019). Predicting long-term publication impact through a combination of early citations and journal impact factor. *Journal of Informetrics,* 13(1), 32-49.

Adams, J. (2005). Early citation counts correlate with accumulated impact. *Scientometrics*, 63(3), 567-581.

Anfossi, A., Ciolfi, A., Costa, F., Parisi, G., & Benedetto, S. (2016). Large-scale assessment of research outputs through a weighted combination of bibliometric indicators. *Scientometrics, 107*(2), 671-683.

Baumgartner, S., & Leydesdorff, L. (2014). Group-based trajectory modelling (GBTM) of citations in scholarly literature: Dynamic qualities of "transient" and "sticky" knowledge claims. *Journal of the American Society for Information Science and Technology*, 65(4), 797-811.

Bloor, D. (1976). *Knowledge and Social Imagery*. London: Routledge, Kegan and Paul.

Bornmann, L., & Daniel, H.D. (2008). What do citation counts measure? A review of studies on citing behavior. *Journal of Documentation*, 64(1), 45–80.

Bornmann, L., Leydesdorff, L., & Wang, J. (2014). How to improve the prediction based on citation impact percentiles for years shortly after the publication date? *Journal of Informetrics*, 8(1), 175-180.

Garfield, E. (1972). Citation analysis as a tool in journal evaluation. *Science,* 178, 471-479.

Gilbert, G.N., (1977). Referencing as persuasion. *Social Studies of Science,* 7(1), 113–122.

Glänzel, W., Schlemmer, B., & Thijs, B. (2003). Better late than never? On the chance to become highly cited only beyond the standard bibliometric time horizon. *Scientometrics, 58*(3), 571-586.

Latour, B., (1987). Science in action: How to follow scientists and engineers through society. Cambridge, MA*: Harvard University Press.*

Levitt, J. M., & Thelwall, M. (2011). A combined bibliometric indicator to predict article impact. *Information Processing and Management, 47*(2), 300-308.

MacRoberts, M. H., & MacRoberts, B. R. (2018). The mismeasure of science: Citation analysis. *Journal of the Association for Information Science and Technology,* 69(3), 474-482.

MacRoberts, M. H., & MacRoberts, B. R., (1984). The negational reference: Or the art of dissembling. *Social Studies of Science,* 14(1), 91–94.

MacRoberts, M. H., & MacRoberts, B. R., (1987). Another test of the normative theory of citing. *Journal of the American Society for Information Science,* 38(4), 305–306.

MacRoberts, M. H., & MacRoberts, B. R., (1988). Author motivation for not citing influences: A methodological note. *Journal of the American Society for Information Science,* 39(6), 432–433.

MacRoberts, M. H., & MacRoberts, B. R., (1989a). Citation analysis and the science policy arena. *Trends in Biochemical Science,* 14(1), 8–10.

MacRoberts, M. H., & MacRoberts, B. R., (1989b). Problems of citation analysis: A critical review. *Journal of the American Society for Information Science,* 40(5), 342–349.

MacRoberts, M. H., & MacRoberts, B. R., (1996). Problems of citation analysis. *Scientometrics,* 36(3), 435–444.

Mingers, J. (2008). Exploring the dynamics of journal citations: modelling with S-curves. *Journal Operational Research Society*, 59 (8), 1013-1025.

Mulkay, M. (1976). Norms and ideology in science. *Social Science Information,* 15 (4-5), 637-656.

Onodera, N. (2016). Properties of an index of citation durability of an article. *Journal of Informetrics, 10*(4), 981-1004.

Rousseau, R. (1988). Citation distribution of pure mathematics journals. In: Egghe, L., Rousseau, R. (Ed.) Informetrics, Belgium: Diepenbeek, 87/88, 249-262, *Proceedings 1st International Conference on Bibliometrics and Theoretical Aspects of Information Retrieval*.

Song, Y., Situ, F. L., Zhu, H. J., & Lei, J. Z. (2018). To be the Prince to wake up Sleeping Beauty: The rediscovery of the delayed recognition studies. *Scientometrics, 117*(1), 9–24.

Stegehuis, C., Litvak, N., & Waltman, L. (2015). Predicting the long-term citation impact of recent publications. *Journal of Informetrics,* 9(3), 642-657.

Stern, D.I. (2014). High-ranked social science journal articles can be identified from early citation information. *PLoS One*, 9(11), 1-11.

Stringer, M.J., Sales-Pardo, M., & Amaral, L.A.N. (2008). Effectiveness of journal ranking schemes as a tool for locating information. *PLoS ONE,* 3(2) doi:10.1371/journal.pone.0001683.

Teixeira, A. A. C., Vieira, P. C., & Abreu, A. P. (2017). Sleeping Beauties and their princes in innovation studies. *Scientometrics, 110*(2), 541–580.).

Teplitskiy, M., Dueder, E., Menietti, M., & Lakhani, K. (2019). Why citations don't mean what we think they mean: Evidence from citers. Proceedings of the *17th International Society of Scientometrics and Informetrics Conference - (ISSI 2019)* 2-5 September 2019, Rome, Italy.

van Raan, A.F.J. (2004). Sleeping beauties in science. *Scientometrics*, 59(3), 461–466.

Wang, J. (2013). Citation time window choice for research impact evaluation. *Scientometrics,* 94(3), 851-872.